# Modelling Vaporous Cavitation on Fluid Transients

Jian-Jun SHU

*School of Mechanical & Aerospace Engineering,*

*Nanyang Technological University, 50 Nanyang Avenue, Singapore 639798*

## ABSTRACT

A comprehensive study of the problem of modelling *vaporous cavitation* in transmission lines is presented. The *two-phase homogeneous equilibrium vaporous cavitation model* which has been developed is compared with the conventional *column separation model*. The latter predicts unrealistically high pressure spikes because of a conflict arising from the prediction of negative cavity sizes if the pressure is not permitted to fall below the vapour pressure, or the prediction of negative absolute pressures if the cavity size remains positive. This is verified by a comparison of predictions with previously published experimental results on upstream, midstream and downstream cavitation. The new model has been extended to include frequency-dependent friction. The characteristics predicted by the *frequency-dependent friction model* show close correspondence with experimental data.

# 1   INTRODUCTION

In hydraulic systems, cavitation can have a serious effect upon the performance of pumps, valves and other components. For the case of positive displacement pumps, the collapse of cavitation bubbles may cause surface damage to the pump and to other components as the surface debris is carried through the system by the working fluid. Severe cavitation reduces volumetric efficiency. Consequently, in order to improve the performance and reliability of systems, it is important to be able to predict the onset and degree of cavitation taking place. This will enable improvements to be made to both pump and circuit design.

Fluid mixtures in hydraulic systems can be classified into five states according to the working conditions: (1) fully degassed liquid; (2) fully degassed liquid with vapour; (3) liquid with dissolved gas; (4) liquid with dissolved and undissolved gas and (5) liquid with dissolved gas, undissolved gas and vapour. If we consider a liquid with dissolved gas, then depending upon the magnitude of the pressure reduction which is occurring in the system and its rate of change, gas cavities may initially be formed due to the presence of suitable nuclei (contaminant, for example). These cavities grow slowly by the diffusion of dissolved gas evolving from the liquid. The process is called *gaseous cavitation*[1,2]. When the pressure falls below the vapour pressure of a liquid, the cavities grow very rapidly because of evaporation into the growing cavity. The process is called *vaporous cavitation* which is, in a sense, a true cavitation. At present the general approach in transmission line modelling is to assume column separation (see, for example, Wylie and Streeter[3,4], Simpson and Wylie[5], and Bergant and Simpson[6]). However, such an approach is over-simplistic and can lead to unrealistic predictions. In this paper a brief review of the classical approach to *vaporous cavitation* is presented and followed by the development of an improved model. When vaporous cavities are locally incipient or desinent, the local pressures may be less than or greater than the vapour pressure of the liquid. However, for modelling purposes in engineering, it is assumed that the local pressures are equal to the vapour pressure when *vaporous cavitation* is taking place.

# 2 COLUMN SEPARATION

For the cases where the dissolved content of a liquid is low (for example, water at atmospheric pressure and $25^oC$ contains 1:84% dissolved air by volume) cavities are primarily in the form of *vaporous cavitation* when the vapour pressure of the liquid is reached.

An early attempt to evaluate the effect of *vaporous cavitation* has been described by Siemons[7], followed by Baltzer[8]. Both assumed that in the top part of a horizontal tube, a thin cavity develops containing vapour at a constant vapour pressure. The flow underneath can be considered incompressible, which enables a mathematical model to be established. The results do not have general validity because the test case is arbitrary.

Without focusing on the bubble dynamics of two-phase flow, Wylie and Streeter[3,4] divided a tube into $N$ equal elements, each $\Delta x$ in length as shown in Figure 1. The vaporous cavities are assumed to be concentrated at fixed computing nodes and no cavities exist within each element. The one-dimensional equations of continuity and momentum are

$$\frac{1}{c_0^2}\frac{\partial P}{\partial t} + \frac{\rho_l}{\pi r_0^2}\frac{\partial Q}{\partial x} = 0 \qquad (1)$$

$$\frac{\rho_l}{\pi r_0^2}\frac{\partial Q}{\partial t} + \frac{\partial P}{\partial x} + F_0(Q) + \rho_l g \sin\theta_0 = 0 \qquad (2)$$

where $P$, $Q$ are instantaneous pressure and volumetric flow rate of the liquid, which has liquid density $\rho_l$; $r_0$ is the internal radius of tube; $\theta_0$ is the inclination of the tube to the horizontal which is positive when the elevation increases in the $+x$ direction, measured along the tube; $c_0$ is the speed of sound in the tube:

$$c_0^2 = \frac{K_l}{\rho_l}.$$

When the elasticity of the tube wall is considered, $c_0$ will be modified as

$$c_0^2 = \frac{K_l/\rho_l}{1 + \frac{2K_l r_0}{Ee}},$$

where $K_l$, $E$ and $e$ are bulk modulus of elasticity of the liquid, Young's modulus and thickness of the tube respectively.

For laminar flow, the viscous loss term $F_0(Q)$ can be expressed in terms of the one-dimensional linear resistance compressible flow model[9–12], or the 'exact' first-order model[13–19], or other distributed parameter approaches[20;21]; For turbulent flow, a model with the Darcy-Weisbach friction factor $f$[3;4] is used

$$F_0(Q) = \frac{f\rho_l Q|Q|}{4\pi^2 r_0^5}. \tag{3}$$

Using the method of characteristics, two finite-difference equations are expressed as

$$C^+ : \frac{\rho_l}{\pi r_0^2} Q_D + \frac{1}{c_0} P_D = C_A \tag{4}$$

$$C^- : \frac{\rho_l}{\pi r_0^2} Q_D - \frac{1}{c_0} P_D = C_B \tag{5}$$

in which $C_A$ and $C_B$ are always known constants when the equations are applied:

$$C_A = \frac{\rho_l}{\pi r_0^2} Q_A + \frac{1}{c_0} P_A - \frac{\Delta x}{c_0} F_0(Q_A) - \frac{\rho_l g \Delta x}{c_0} \sin\theta_0 \tag{6}$$

$$C_B = \frac{\rho_l}{\pi r_0^2} Q_B - \frac{1}{c_0} P_B - \frac{\Delta x}{c_0} F_0(Q_B) - \frac{\rho_l g \Delta x}{c_0} \sin\theta_0. \tag{7}$$

Ignoring mass transfer during cavitation, a continuity equation for the vapour volume $V$ is applied at each computing node

$$\frac{\partial V}{\partial t} = Q_2 - Q_1. \tag{8}$$

Two subscripts 1 and 2 are used to indicate left and right limits at each computing node. The resultant finite-difference equation is

$$V_D = V_E + \frac{\Delta x}{2c_0}(Q_{D_2} - Q_{D_1} + Q_{E_2} - Q_{E_1}). \tag{9}$$

This model is based on the *column separation* hypothesis that the flow of liquid in the tube is instantaneously and completely separated by its vapour phase when the cavity is formed. For the magnitudes of the transient pressures and velocities ordinarily encountered with viscous flow in a horizontal or near-horizontal tube terminated by a valve, the *column separation* hypothesis does not imply complete physical interruption of the flowing liquid. In fact, once development of a cavity has occurred at some point in the pipe, the cavity usually expands and propagates in the direction of flow as an elongated bubble. So *column separation* does not necessarily occur in practice, especially in a horizontal or near-horizontal tube.

During the existence of the cavity, an internal boundary condition is established in the *vapour column separation model* at each computing node. A computer program chart to solve the problem of *vaporous cavitation* is presented by Wylie and Streeter[3,4] and is summarized in Appendix 1.

# 3 TWO-PHASE HOMOGENEOUS EQUILIBRIUM VAPOROUS CAVITATION

Although the *vapour column separation model* is easily implemented and faithfully reproduces many of the essential features of a physical event, it has some serious deficiencies:

(i) To avoid the prediction of a negative cavity size (if the pressure is not permitted to fall below the vapour pressure), or the prediction of negative absolute pressures (if the cavity size remains positive), artificial restrictions are imposed (see Figure 2 and the steps 7 and 9 in Appendix 1). These result in unrealistically large pressure spikes that discredit the overall value of the numerical results.

(ii) The internal boundary condition (9) permits, subjectively, vapour cavities to be formed only at computing nodes. The simulation results are strongly biased according to where the computing nodes are located.

(iii) Because the size of the cavity and its mass transfer are ignored at each computing node, this model is clearly limited in its ability to model cavitation correctly.

(iv) At each computing node, a flow rate discontinuity is assumed. Hence, for a given location in the pipe, there will be two predicted values of flow rate, which is clearly inconsistent with the true (measured) behaviour at this point. In addition, the difference between the two predicted values increases when there is a high degree of cavitation and also when the number of computing nodes is small. However, when a large number of computing nodes is used, there are a corresponding number of discontinuities leading to a mathematical model that is ill defined.

When the pressure falls to the vapour pressure of the liquid during transient flows, vaporization occurs. Vapour cavities may be physically dispersed homogeneously (bubbles evenly distributed), or collected into a single or multiple void space, or a combination of the two. The behaviour can be described by two-phase flow theory. Single-phase flow becomes a special case when no vaporization occurs. In this section a *two-* *phase homogeneous equilibrium vaporous cavitation model* is presented as an alternative formulation to overcome the above deficiencies and improve the reliability of numerical modelling of *vaporous cavitation*.

A difference in velocity between the liquid phase and vapour phase will promote mutual momentum transfer. Often these processes proceed very rapidly and it can be assumed that equilibrium conditions prevail. In other words, a vapour cavity shares the same velocity and pressure as the liquid flow at the point where *vaporous cavitation* occurs. The average values of velocity in the two-phase mixture are the same as the values for each phase and, in this paper, the author proposes a mathematical model named *homogeneous equilibrium flow model*.

The basic equations for the unsteady *homogeneous equilibrium flow model* in a tube are:

$$\frac{1}{c_0^2}\frac{\partial P}{\partial t} + (\rho_l - \rho_v)\frac{\partial \alpha}{\partial t} + \frac{\rho_m}{\pi r_0^2}\frac{\partial}{\partial x}(\frac{Q}{\alpha}) = 0 \tag{10}$$

$$\frac{\rho_m}{\pi r_0^2}\frac{\partial}{\partial t}(\frac{Q}{\alpha}) + \frac{\partial P}{\partial x} + F_0(\frac{Q}{\alpha}, \alpha) + \rho_m g \sin\theta_0 = 0. \tag{11}$$

In terms of the volumetric fraction $\alpha$ of liquid and vapour phase density $\rho_v$, the mean density $\rho_m$ can be expressed in

$$\rho_m = \alpha \rho_l + (1-\alpha)\rho_v.$$

In the above equations the second term in (10) describes the interfacial mass transfer rate and the term $Q/\alpha$ in (10) and (11) indicates the difference between the liquid phase flowrate and the vapour phase flowrate. Using the Darcy-Weisbach friction factor $f$, the term $F_0(Q/\alpha, \alpha)$ can be expressed as

$$F_0(\frac{Q}{\alpha}, \alpha) = \frac{f\rho_m Q|Q|}{4\pi^2 \alpha^2 r_0^5}. \tag{12}$$

The method of characteristics is used to transform the above equations to four ordinary differential equations

$$\left.\begin{array}{l} \frac{1}{\pi r_0^2}\frac{d}{dt}(\frac{Q}{\alpha}) + \frac{1}{\rho_l c_0}\frac{d}{dt}(P - p_v) + c_0\frac{\partial}{\partial t}(\ln\frac{\rho_m}{\rho_l}) + \frac{fQ|Q|}{4\pi^2 \alpha^2 r_0^5} + g\sin\theta_0 = 0 \\ \frac{dx}{dt} = c_0 \end{array}\right\} C^+$$

$$\left.\begin{array}{l} \frac{1}{\pi r_0^2}\frac{d}{dt}(\frac{Q}{\alpha}) - \frac{1}{\rho_l c_0}\frac{d}{dt}(P - p_v) - c_0\frac{\partial}{\partial t}(\ln\frac{\rho_m}{\rho_l}) + \frac{fQ|Q|}{4\pi^2 \alpha^2 r_0^5} + g\sin\theta_0 = 0 \\ \frac{dx}{dt} = -c_0 \end{array}\right\} C^-.$$

The equations needed to solve for the variables at each time step are:

$$C^+ : \frac{1}{\pi r_0^2} \frac{Q_D}{\alpha_D} + \frac{1}{\rho_l c_0}(P_D - p_v) + \frac{c_0}{2} \ln \frac{\rho_{m_D}}{\rho_l} = C_A \qquad (13)$$

$$C^- : \frac{1}{\pi r_0^2} \frac{Q_D}{\alpha_D} - \frac{1}{\rho_l c_0}(P_D - p_v) - \frac{c_0}{2} \ln \frac{\rho_{m_D}}{\rho_l} = C_B. \qquad (14)$$

By solving for $Q_D$, $P_D$ and $\alpha_D = (\rho_{m_D} - \rho_v)/(\rho_l - \rho_v)$, $C_A$ and $C_B$ are known constants when the equations are applied:

$$C_A = \frac{1}{\pi r_0^2} \frac{Q_A}{\alpha_A} + \frac{1}{\rho_l c_0}(P_A - p_v) + \frac{c_0}{2} \ln \frac{\rho_{m_E}\rho_{m_F}}{\rho_l \rho_{m_A}} - \frac{f\Delta x Q_A |Q_A|}{4\pi^2 \alpha_A^2 r_0^5 c_0} - \frac{g\Delta x}{c_0} \sin\theta_0, \qquad (15)$$

$$C_B = \frac{1}{\pi r_0^2} \frac{Q_B}{\alpha_B} - \frac{1}{\rho_l c_0}(P_B - p_v) - \frac{c_0}{2} \ln \frac{\rho_{m_E}\rho_{m_G}}{\rho_l \rho_{m_B}} - \frac{f\Delta x Q_B |Q_B|}{4\pi^2 \alpha_B^2 r_0^5 c_0} - \frac{g\Delta x}{c_0} \sin\theta_0. \qquad (16)$$

Unlike the *vapour column separation model*, no conflict between negative cavity sizes and pressures below the vapour pressure is introduced

$$\text{if} \quad C_A \geq C_B, \quad \text{then} \quad \alpha_D = 1, \quad P_D = \frac{\rho_l c_0}{2}(C_A - C_B) + p_v; \qquad (17)$$

$$\text{if} \quad C_A < C_B, \quad \text{then} \quad P_D = p_v, \quad \alpha_D = \frac{\rho_l \exp \frac{C_A - C_B}{c_0} - \rho_v}{\rho_l - \rho_v}. \qquad (18)$$

In either case,

$$Q_D = \frac{\pi r_0^2 \alpha_D}{2}(C_A + C_B). \qquad (19)$$

A computer program chart of the *two-phase homogeneous equilibrium vaporous cavitation model* is given in Appendix 2. This is readily implemented and overcomes the major deficiencies of the *vapour column separation model*.

# 4  FREQUENCY-DEPENDENT FRICTION

It is well known that fluid friction losses are dependent not only upon the average velocity at the instant, but also upon the history of the average velocity in transient flow. This is called *frequency-dependent friction* and results in higher frequencies being attenuated much more rapidly than low frequency components of velocity. The Darcy-Weisbach equation (12) does not account for such effects.

Zielke[22] has developed appropriate friction terms using a weighting function for transient laminar flow, but direct numerical integration of the equations requires a large amount of calculation. Some efficient procedures have been developed by Trikha[23] and then Kagawa *et al.*[24]. The procedures consist of approximating the exact Zielke weighting function as a sum of impulse responses of first order lags. Much less computer storage and computation time are two important benefits. The error incurred as a result of the approximation is adjustable using the method posed by Kagawa *et al.*[24] and is therefore much more accurate than Trikha model. Consequently, in this paper, Kagawa's weighting function will be adopted.

Good progress has been made in developing relations for turbulent flow with *frequency-dependent friction* by Hirose[25], Margolis and Brown[26], Brown[27,28], Brown and Linney[29], Vardy and Brown[30], and Shu[31]. However, little work appears to have been undertaken on *frequency-dependent friction* under two-phase flow conditions. Reasoning from an analogy to laminar flow, the author suggests using the following expression for *frequency-dependent friction* $F(Q/\alpha, \alpha)$ in turbulent, two-phase flow

$$F(\frac{Q}{\alpha}, \alpha) = F_0(\frac{Q}{\alpha}, \alpha) + \frac{1}{2}(W * \frac{\partial F_0}{\partial t})(t) \tag{20}$$

where the convolution integral is defined as $(f * g)(t) = \int_0^t f(t-\lambda)g(\lambda)d\lambda$. $F_0(Q/\alpha, \alpha)$ is a steady friction term and $W(t)$ is a weighting function. The foregoing relationship is based upon the following:

(i) The nature of the weighting function is to smooth rapid changes in flow velocity for laminar flow conditions. Here the author proposes a more general smoothing function that may be applied to laminar, turbulent and multi-phase flow.

(ii) Viscosity effects are not well defined in turbulent two-phase flows. No viscosity appears explicitly in the relationship (20).

(iii) The well-defined case of transient laminar flow is inclusive.

(iv) The relationship has sufficient generality that the weighting function can be designed by some given criteria depending on the application; the manner is analogous to smoothing filter design in signal processing.

The steady-state friction value $F_0(Q/\alpha, \alpha)$ is the same as (12)

$$F_0(\frac{Q}{\alpha}, \alpha) = \frac{f\rho_m Q|Q|}{4\pi^2 \alpha^2 r_0^5}. \tag{21}$$

The weighting function $W(t)$ can be approximated by sum of $k$ impulse responses of first order elements.

$$W(t) = \sum_{i=1}^{k} m_i e^{-n_i \tau} \tag{22}$$

where $m_i$ and $n_i$ can be determined by curve fitting[24] to the Zielke weighting function as listed in Table 1. The parameter $\tau = \mu_m t/\rho_m r_0^2$ is a non-dimensional delay time and the "virtual" viscosity $\mu_m$ for the two-phase mixture is expressed by Dukler's expression[32;33] in terms of liquid viscosity $\mu_l$ and vapour viscosity $\mu_v$

$$\mu_m = \alpha\mu_l + (1-\alpha)\mu_v. \tag{23}$$

A simple recursive form for $F(Q/\alpha, \alpha)$ in equation (20) has been developed by Kagawa et al.[24] that only requires characteristic data from the last time step.

$$F|_{t+\Delta t} = F_0|_{t+\Delta t} + \frac{1}{2}\sum_{i=1}^{k} y_i(t+\Delta t) \tag{24}$$

with

$$\begin{cases} y_i(t+\Delta t) = y_i(t)e^{-n_i\Delta\tau} + m_i e^{-n_i\frac{\Delta\tau}{2}}[F_0|_{t+\Delta t} - F_0|_t] \\ y_i(0) = 0 \end{cases} \tag{25}$$

where

$$\Delta\tau = \frac{\Delta t}{r_0^2}\frac{\mu_m}{\rho_m}\Big|_{t+\frac{\Delta t}{2}}. \tag{26}$$

From (25), it is evident that a good approximation to $W(t)$ over a region of $\tau > \Delta\tau/2$ makes the numerical integration approach presented by Zielke accurate. In Table 1, $\tau_{m_i}$ is such $\tau$ that the relative error between $W(t)$ and the Zielke weighting function[22] becomes less than 1% over a region of $\tau > \tau_{m_i}$ and for a calculation error of less than 1%, $k$ is determined according to $\Delta\tau/2 > \tau_{m_i}$. Using the characteristics method for equations (10), (11) together with characteristics equations (13), (14), the following is obtained,

$$\Delta\tau_A = \frac{\Delta x}{2r_0^2 c_0}\left(\frac{\mu_{m_A}}{\rho_{m_A}} + \frac{\mu_{m_F}}{\rho_{m_F}}\right), \quad \Delta\tau_B = \frac{\Delta x}{2r_0^2 c_0}\left(\frac{\mu_{m_B}}{\rho_{m_B}} + \frac{\mu_{m_G}}{\rho_{m_G}}\right) \tag{27}$$

and

$$F_A = \frac{f\rho_{m_A}Q_A|Q_A|}{4\pi^2 \alpha_A^2 r_0^5} + \frac{1}{2}\sum_{i=1}^{k}[y_{iA}e^{-n_i\Delta\tau_A} + \frac{fm_i}{4\pi^2 r_0^5}e^{-n_i\frac{\Delta\tau_A}{2}}\left(\frac{\rho_{m_A}Q_A|Q_A|}{\alpha_A^2} - \frac{\rho_{m_F}Q_F|Q_F|}{\alpha_F^2}\right)], \tag{28}$$

$$F_B = \frac{f\rho_{m_B}Q_B|Q_B|}{4\pi^2 \alpha_B^2 r_0^5} + \frac{1}{2}\sum_{i=1}^{k}[y_{iB}e^{-n_i\Delta\tau_B} + \frac{fm_i}{4\pi^2 r_0^5}e^{-n_i\frac{\Delta\tau_B}{2}}\left(\frac{\rho_{m_B}Q_B|Q_B|}{\alpha_B^2} - \frac{\rho_{m_G}Q_G|Q_G|}{\alpha_G^2}\right)] \tag{29}$$

and

$$C_A = \frac{1}{\pi r_0^2}\frac{Q_A}{\alpha_A} + \frac{1}{\rho_l c_0}(P_A - p_v) + \frac{c_0}{2}\ln\frac{\rho_{m_E}\rho_{m_F}}{\rho_l\rho_{m_A}} - \frac{\Delta x F_A}{\rho_{m_A}c_0} - \frac{g\Delta x}{c_0}\sin\theta_0, \tag{30}$$

$$C_B = \frac{1}{\pi r_0^2} \frac{Q_B}{\alpha_B} - \frac{1}{\rho_l c_0}(P_B - p_v) - \frac{c_0}{2} \ln \frac{\rho_{m_E} \rho_{m_G}}{\rho_l \rho_{m_B}} - \frac{\Delta x F_B}{\rho_{m_B} c_0} - \frac{g \Delta x}{c_0} \sin \theta_0. \qquad (31)$$

The computer program chart of the *two-phase homogeneous equilibrium vaporous cavitation model* with *frequency-dependent friction* is presented in Appendix 3.

# 5 EXPERIMENTAL VERIFICATION

Experiments on *vaporous cavitation* have been conducted by Sanada et al.[34] using a horizontal acrylic pipeline and water as the working fluid. They examined *vaporous cavitation* which they classified as upstream-, midstream- and downstream-type. The test rig is shown in Figure 3 and the essential parameters are listed in Table 2.

When an upstream valve is rapidly closed in a flowing liquid line, the pressure at the vicinity of the valve is reduced to the vapour pressure and cavities are formed. After a certain time, a reflected pressure wave will collapse the cavities and possibly cause very high pressures at the valve. Several cycles of cavity formation and collapse occur before frictional effects damp the oscillations sufficiently for the minimum pressure at the valve to remain permanently above the vapour pressure. This is *upstream cavitation*. Hydraulic transients in pump delivery lines induced by pump failure[35–38] result in *upstream cavitation*.

When the upstream pressure quickly drops down to a certain level in a flowing liquid line, rarefaction waves will be sent downstream. The pressure at some location in the line may be reduced to the vapour pressure and cavities may be formed. The reflected pressure waves and rarefaction waves in the line will have a dominant effect on the behaviour of the cavities. This is *midstream cavitation*. Hydraulic transients in pump delivery lines induced by pump pulsation[39] may result in *midstream cavitation*.

When the downstream valve is suddenly shut in a flowing liquid line, the liquid will accumulate in the vicinity of the valve. A high pressure at the vicinity of the valve cause flow reversal and several cycles of cavity formation and collapse will consequently occur as a result of fluid accumulation and dissipation. This is *downstream cavitation*. Hydraulic transients in pump suction lines induced by pump failure[8,40] result in *downstream cavitation*.

In the experimental studies conducted by Sanada et al.[34] the absolute pressure head, $H = P/\rho_l g$, was recorded as a function of time at different locations, according to the class of cavitation being investigated.

Figure 4 shows the measured behaviour together with predictions from the various models for upstream cavitation. For this case, where cavitation is severe, the predictions all show similar characteristics, but the high frequency components superimposed on the waveform are rather better damped when frequency-dependent friction is included; this is closer to the measured response. The superiority of the frequency-dependent friction model is much more apparent for the case of midstream cavitation (Figure 5) and downstream cavitation (Figure 6). For both cases, there is good agreement between the frequency-dependent friction model predictions and observed behaviour. This is because the weighting functions in Table 1 correspond to laminar, single-phase flow conditions, which match conditions more closely when cavitation is not too severe.

# 6 CONCLUSIONS

The *Method of Characteristics* approach to the modelling of wave transients in fluid transmission lines has been extended to include two-phase flow following *vaporous cavitation*. The resultant equations, (10), (11) and (20), predict closely the pressure transients for *vaporous cavitation* conditions following sudden closure of a valve in a pipeline; the largest errors are encountered when cavitation is severe. The use of the mathematical model (10), (11) and (20) provides a powerful and basic tool to understand the mechanism of transient fluid flow with *vaporous cavitation* in transmission lines. The modifying weighting function for frequency-dependent friction (20) is a key to producing improved predictions for two-phase transient flows. More work is required to establish appropriate weighting functions particularly for the case of two-phase flow.

# APPENDIX 1: COMPUTER PROGRAM CHART FOR VAPOUR COLUMN SEPARATION MODEL

**Step 1.**   Calculate initial steady-state conditions for $P$, $Q$ and $V$

**Step 2.**   Increment the time by $\Delta t = \Delta x / c_0$

**Step 3.**   If $V_E > 0$, go to step 5

**Step 4.**   Calculate $P_D$ from both (4) and (5) together and if $P_D \geq p_v$ (vapour pressure), go to step 10

**Step 5.**   Set $P_D = p_v$ and calculate $Q_{D_1}$ from (4) and $Q_{D_2}$ from (5) separately

**Step 6.**   Calculate $V_D$ from (9) and if $V_D > 0$, go to step 11

**Step 7.**   Set $V_D = 0$ and calculate $P_D$ from both (4) and (5) together

**Step 8.**   If $P_D \geq p_v$, go to step 10

**Step 9.**   Set $P_D = p_v$ and $V_D = 0$

**Step 10.**  Set $V_D = 0$ and calculate $Q_{D_1} = Q_{D_2}$ from both (4) and (5) together

**Step 11.**  Calculate boundary conditions for $P$, $Q$ and $V$

**Step 12.**  Check to see if a given maximum time tolerance has been exceeded. If not, go to step 2; Otherwise, terminate calculations and print out results.

# APPENDIX 2: COMPUTER PROGRAM CHART FOR TWO-PHASE HOMOGENEOUS EQUILIBRIUM VAPOROUS CAVITATION MODEL

**Step 1.** Calculate initial steady-state conditions for $P$, $Q$ and

**Step 2.** Increment the time by $\Delta t = \Delta x / c_0$

**Step 3.** Calculate $C_A$ from (15) and $C_B$ from (16)

**Step 4.** If $C_A \geq C_B$, set $\alpha_D = 1$ and calculate $P_D$ from (17)

**Step 5.** If $C_A < C_B$, set $P_D = p_v$ and calculate $\alpha_D$ from (18)

**Step 6.** Calculate $Q_D$ from (19)

**Step 7.** Calculate boundary conditions for $P$, $Q$ and $\alpha$

**Step 8.** Check to see if a given maximum time tolerance has been exceeded. If not, go to step 2; otherwise terminate the calculations and print out the results.

# APPENDIX 3: COMPUTER PROGRAM CHART FOR TWO-PHASE HOMOGENEOUS EQUILIBRIUM VAPOROUS CAVITATION MODEL WITH FREQUENCY-DEPENDENT FRICTION

**Step 1.**   Calculate initial steady-state conditions for $P$, $Q$, $\alpha$ and set $y_i = 0; 1 \leq i \leq 10$ at each computing node

**Step 2.**   Increment the time by $\Delta t = \Delta x/c_0$

**Step 3.**   Calculate $\Delta \tau_A$ and $\Delta \tau_B$ from (27) and determining $k$ satisfied by $\min(\Delta \tau_A/2; \Delta \tau_B/2) > \tau_{m_i}$ from the Table 1

**Step 4.**   Calculate $F_A$ from (28) and $F_B$ from (29)

**Step 5.**   Calculate $C_A$ from (30) and $C_B$ from (31)

**Step 6.**   If $C_A \geq C_B$, set $\alpha_D = 1$ and calculate $P_D$ from (17)

**Step 7.**   If $C_A < C_B$, set $P_D = p_v$ and calculate $\alpha_D$ from (18)

**Step 8.**   Calculate $Q_D$ from (19)

**Step 9.**   Calculate boundary conditions for $P$, $Q$ and $\alpha$

**Step 10.**  Updating $y_i; 1 \leq i \leq 10$ from (25)

**Step 11.**  Check to see if a given maximum time tolerance has been exceeded. If not, go to step 2; otherwise, the calculations are terminated and the results are printed out.

# Table Captions

**Table 1:** $n_i$, $m_i$ and $\tau_{m_i}$

**Table 2: Parameter list**

| $i$ | 1 | 2 | 3 | 4 | 5 | 6 | 7 | 8 | 9 | 10 |
|---|---|---|---|---|---|---|---|---|---|---|
| $n_i$ | $2.63744 \times 10^1$ | $7.28033 \times 10^1$ | $1.87424 \times 10^2$ | $5.36626 \times 10^2$ | $1.57060 \times 10^3$ | $4.61813 \times 10^3$ | $1.36011 \times 10^4$ | $4.00825 \times 10^4$ | $1.18153 \times 10^5$ | $3.48316 \times 10^5$ |
| $m_i$ | $1.0$ | $1.16725$ | $2.20064$ | $3.92861$ | $6.78788$ | $1.16761 \times 10^1$ | $2.00612 \times 10^1$ | $3.44541 \times 10^1$ | $5.91642 \times 10^1$ | $1.01590 \times 10^2$ |
| $\tau_{m_i}$ | $6.2 \times 10^{-2}$ | $2.8 \times 10^{-2}$ | $9.9 \times 10^{-3}$ | $3.3 \times 10^{-3}$ | $1.1 \times 10^{-3}$ | $3.6 \times 10^{-4}$ | $1.2 \times 10^{-4}$ | $4.1 \times 10^{-5}$ | $1.4 \times 10^{-5}$ | $4.7 \times 10^{-6}$ |

Table 1: $n_i$, $m_i$ and $\tau_{m_i}$

|                              | Upstream type | Midstream type | Downstream type |
|------------------------------|---------------|----------------|-----------------|
| Time tolerance ($s$)         | 13.5          | 3.0            | 5.0             |
| Upstream pressure ($bar$)    | 5.49164       | 5.58971        | 4.90325         |
| Downstream pressure ($bar$)  | 0.98065       | 0.98065        | 0.98065         |
| Initial velocity ($m/s$)     | 1.5           | 1.5            | 1.45            |
| Radius $r_0(mm)$             | 7.6           |                |                 |
| Length $L(km)$               | 0.2           |                |                 |
| Liquid density $\rho_l(kg/m^3)$ | 1000       |                |                 |
| Vapour density $\rho_v(kg/m^3)$ | 0.8        |                |                 |
| Acoustic velocity $c_0(m/s)$ | 820           |                |                 |
| Vapour pressure $p_v(bar)$   | 0.023         |                |                 |
| Liquid viscosity $\mu_l(cP)$ | 1.0           |                |                 |
| Vapour viscosity $\mu_v(cP)$ | 0.0087        |                |                 |

Table 2: Parameter list

# Figure Captions

**Figure 1: Numerical scheme**

**Figure 2: Flow chart for** *vapour column separation model*

**Figure 3: Classification of** *vaporous cavitation*

**Figure 4: Upstream type cavitation: (a) Experimental result (Sanada** *et al.*[34]**), (b) Column separation model, (c) Two-phase homogeneous equilibrium vaporous cavitation model, (d) Frequency-dependent friction model**

**Figure 5: Midstream type cavitation: (a) Experimental result (Sanada** *et al.*[34]**), (b) Column separation model, (c) Two-phase homogeneous equilibrium vaporous cavitation model, (d) Frequency-dependent friction model**

**Figure 6: Downstream type cavitation: (a) Experimental result (Sanada** *et al.*[34]**), (b) Column separation model, (c) Two-phase homogeneous equilibrium vaporous cavitation model, (d) Frequency-dependent friction model**

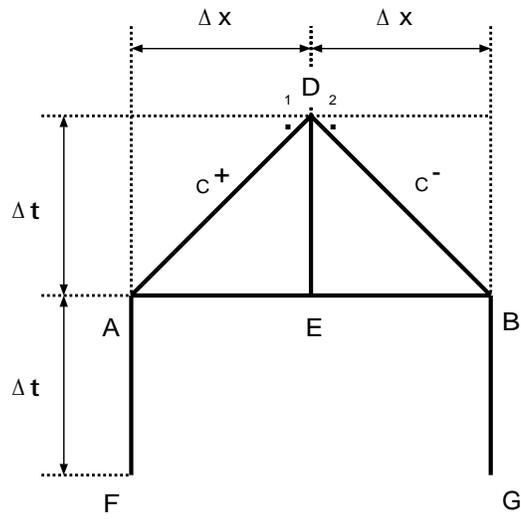

Figure 1: Numerical scheme

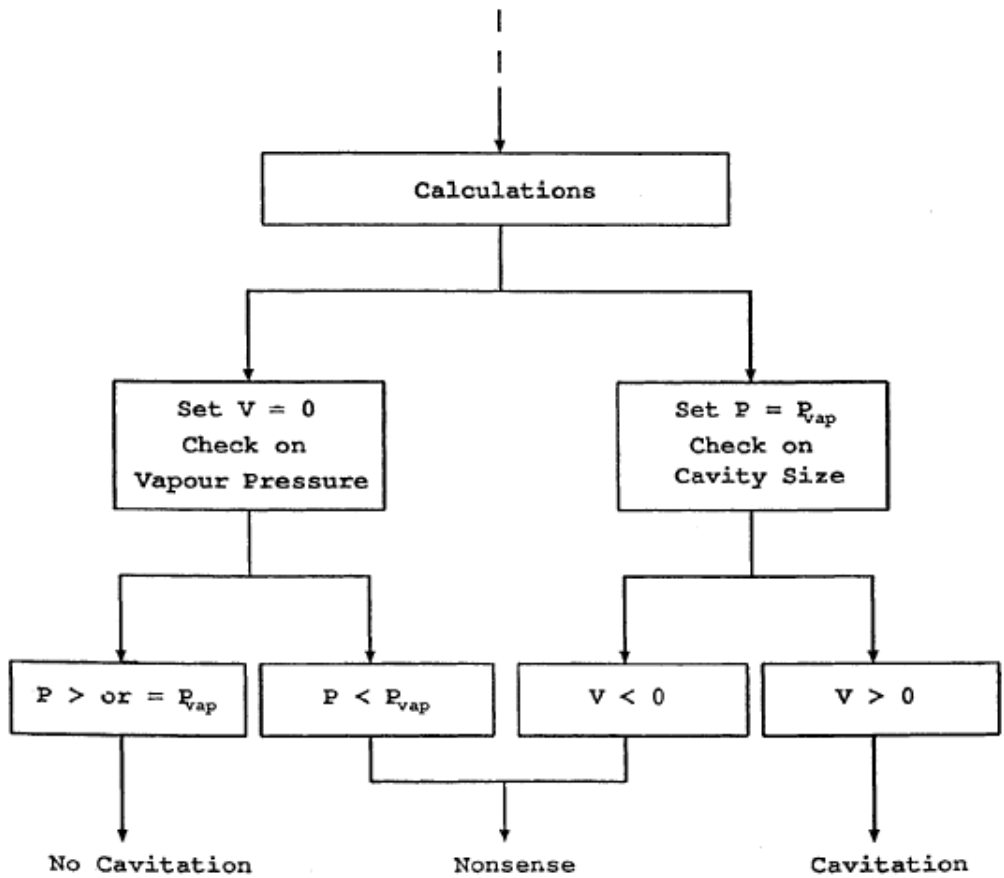

Figure 2: Flow chart for *vapour column separation model*

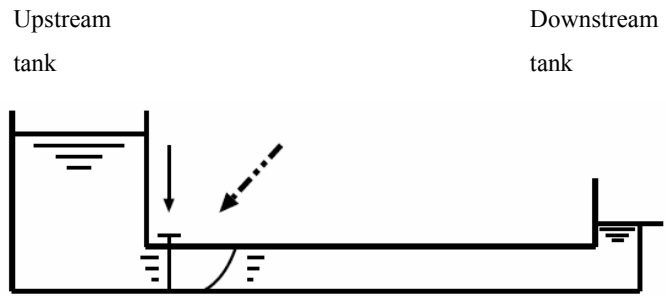

(a) Upstream type

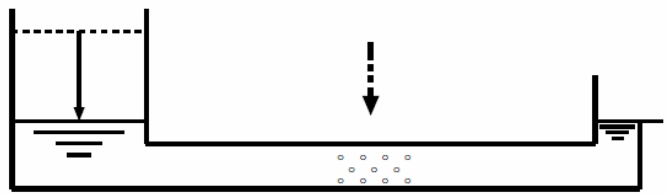

(b) Midstream type

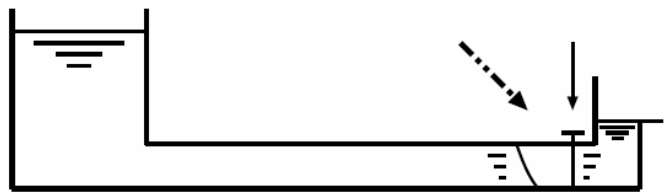

(c) Downstream type

▶ : Cavitation

⟶ : Cause of fluid transients

Figure 3: Classification of *vaporous cavitation*

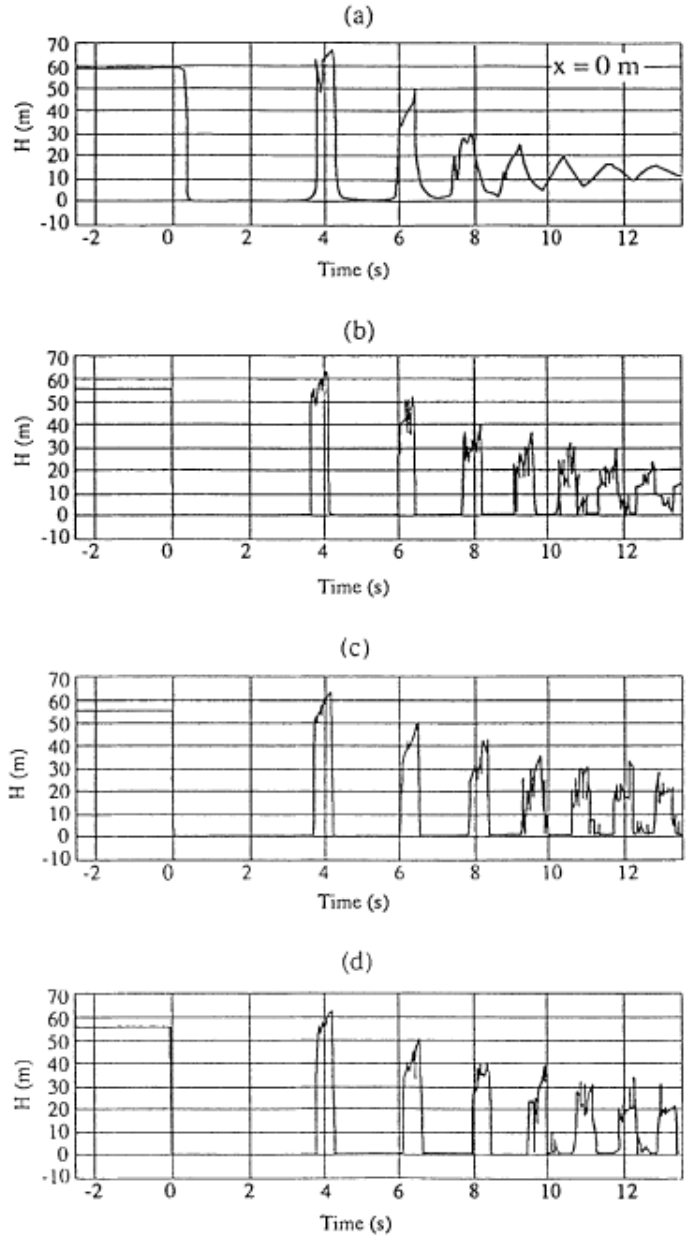

Figure 4: Upstream type cavitation: (a) Experimental result (Sanada et al.[34]), (b) Column separation model, (c) Two-phase homogeneous equilibrium vaporous cavitation model, (d) Frequency-dependent friction model

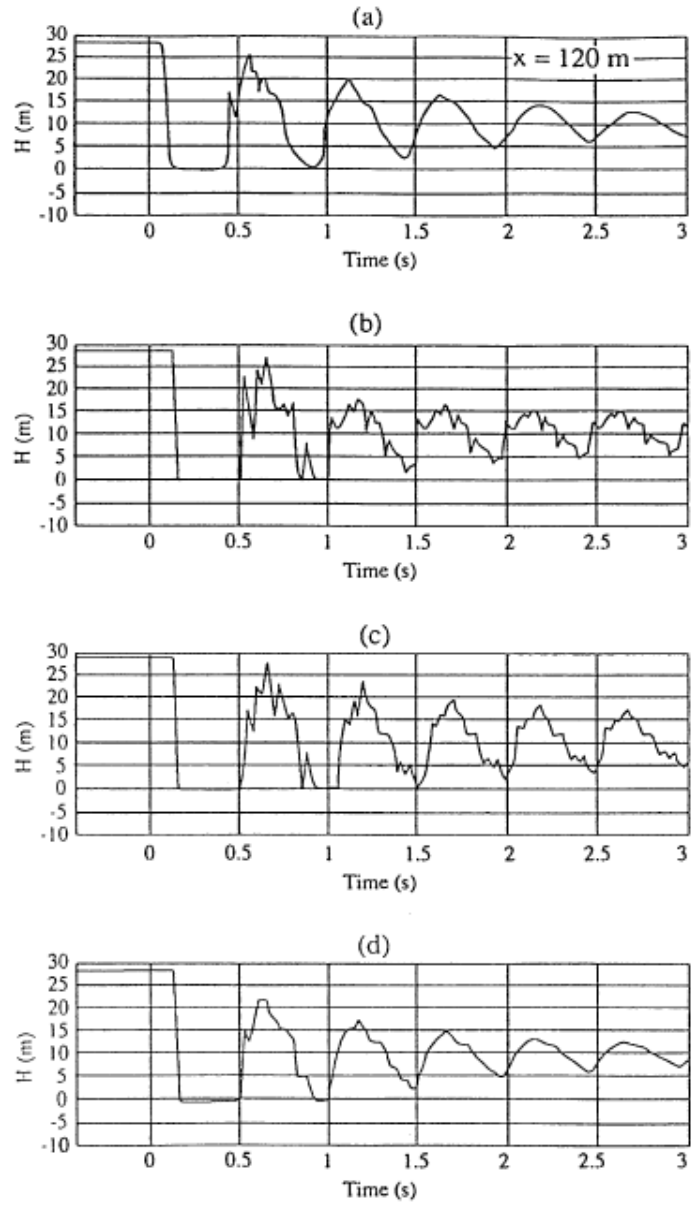

Figure 5: Midstream type cavitation: (a) Experimental result (Sanada *et al.*[34]), (b) Column separation model, (c) Two-phase homogeneous equilibrium vaporous cavitation model, (d) Frequency-dependent friction model

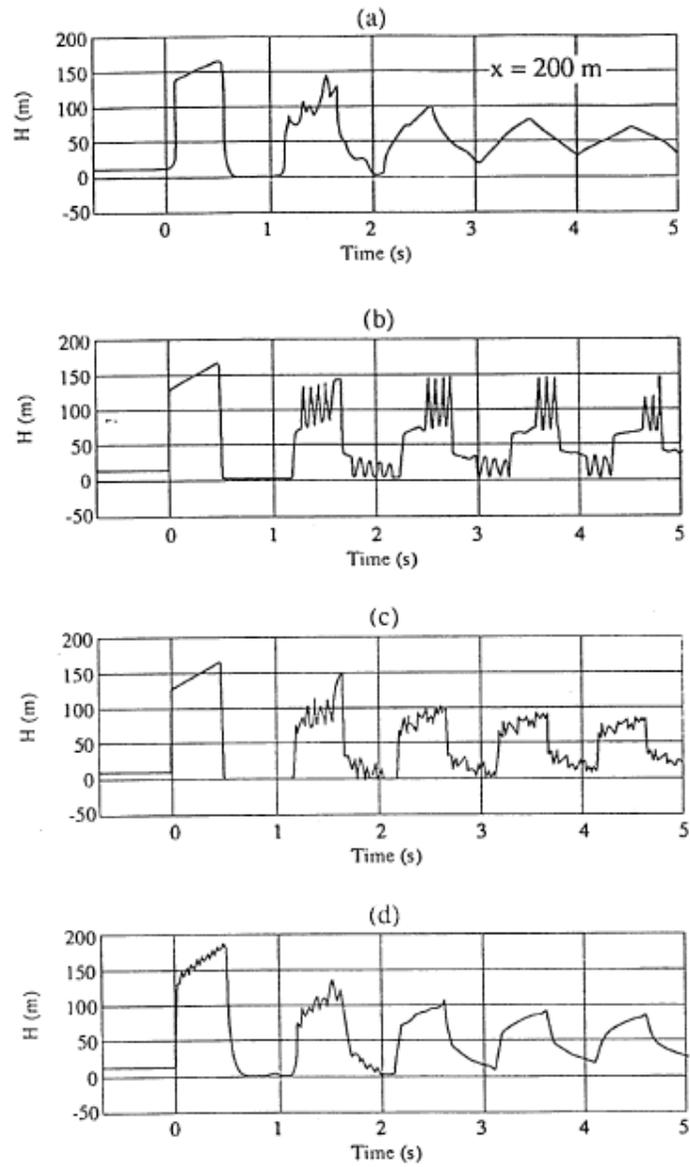

Figure 6: Downstream type cavitation: (a) Experimental result (Sanada *et al.*[34]), (b) Column separation model, (c) Two-phase homogeneous equilibrium vaporous cavitation model, (d) Frequency-dependent friction model

28